# Novel Porous Aluminum Nitride Monolayer Structures: A First-principles Study

Yanwei Luo[1], Jiahui Hu[1], Yu Jia[2,*]

[1] *Department of Physics, College of Science, Henan University of Technology, Zhengzhou 450001, China*

[2] *International Laboratory for Quantum Functional Materials, and School of Physics and Engineering, Zhengzhou University, Zhengzhou 450001, China*

Yanwei Luo (ORCID:0000-0002-5316-9147),

Jiahui Hu (ORCID: 0000-0001-9528-5579),

Yu Jia (ORCID:0000-0002-5514-0639)

## Abstract

By using *ab inito* calculations within Density Functional Theory, we have explored the possible structures and various properties of porous AlN monolayer materials. Two kinds of porous AlN (H- and T-) monolayer are identified, and the phonon dispersion spectrum together with the *ab inito* molecular dynamics simulations demonstrate that their structures are stable. We further show that these AlN (H- and T-) porous monolayers have well-defined porous nanostructures and even higher specific surface areas, namely, 2863 m$^2$/g and 2615 m$^2$/g, which can be comparable with graphene (2630 m$^2$/g), and can also be maintained evenly at high temperatures (>1300K). Furthermore, both porous monolayers exhibit semiconductor properties with 2.89 eV and 2.86 eV indirect band gap, respectively. In addition, the electronic structures of such porous monolayers can be modulated by strains. The band gap of porous T-AlN monolayer experiences an indirect-direct transition when biaxial strain is applied. A moderate

−9% compression can trigger this gap transition. These results indicate that the porous AlN monolayer may potentially be used for optoelectronic applications, as well as for underlying catalysts in the future.

# 1 Introduction

The discovery of varieties of two-dimensional (2D) materials, such as graphene[1], phosphorene, transition metal dichalcogenides as well as tellurene, has attracted lots of attentions, owing to their novel electronic, optical properties and their great potential applications in nano-devices and nanotechnology[2-7] . Lots of new 2D materials with novel structures, as well as electronic and optical properties, have been theoretically explored and experimentally studied along with the development of experimental synthesis and theoretical design technology [8-12]. Among these studies, one interesting finding was that the performance of 2D materials properties could be improved or enhanced dramatically by introducing porous structures [13-18].

As the first discovered 2D monolayer material, porous graphene was the first 2D porous materials to be discovered and studied. In 2009, Bieri and his coworkers fabricated a kind of porous graphene with a regular distribution of the pores [19]. Their studies revealed that the porous graphene exhibits excellent properties which are different from those of graphene due to their unique structure, ie., wide bandgap and high surface areas. Therefore, it was proposed as an ideal material for gas purification and energy storage. A new 2D tetra-symmetrical carbon which possessed similar high Fermi velocity and Dirac-like fermions to graphene was systematically investigated by Liu et al [20]. Meanwhile, another 2D material hexagonal boron nitride (h-BN) was also studied. For instance, two types of porous BN materials were found by using first-principles calculations and global structural search method. The calculations showed that these porous were direct band gap semiconductor and could be applied in the photonic and optoelectronic devices and hydrogen storage [21]. Besides, Wen et al. predicted a novel polymorph of BN with a body-centered tetragonal structure by using ab *inito* calculations [22], and Ma et al predicted the Pnma structure BN [23]. Recently，a series of 2D porous allotropes with new octagonal tiling structure was studied by Li et al[24]. All the 2D allotropes compounded by group V elements could exist stably and possess semiconducting band gaps. Due to their wide band gap from 0.3 to 2.0 eV, these porous materials could be used in the visible and near infrared light optoelectronic devices.

As a new and typical 2D III-nitride material, aluminum nitride (AlN) [25,26] has attracted explosive interests of research due to its many intriguing physical properties such as high chemical stabilities, excellent mechanical properties, high thermal conductivity and significant potential applications [27-30]. Interestingly, the high purity AlN monolayer film has been synthesized by conventional growth methods [31-35]. All these experimental and theoretical progress evoked a significant interest in porous AlN monolayer. AlN monolayer is a typical III-nitride 2D material, with the same number of valence electrons in a unit cell as that of BN. Thus, it is natural to study the similar porous structures of AlN monolayers.

In this paper, two kinds of porous AlN monolayer are explored by employing *ab inito* calculations method within the density functional theory. For each structure, the geometric, energetic and electronic properties have been systematically investigated, with their stabilities analyzed by the phonon dispersion spectrum and molecular dynamics simulations. Furthermore, the relationship between the electronic structure and the in-layer strain of each type is also examined.

## 2 Computational Method

All the first-principles calculations in this work are performed by using Vienna *ab initio* Simulation Package (VASP) [36, 37], which is in the framework of the Density Functional Theory (DFT). The projected augmented wave (PAW) method [38] combined with the generalized gradient approximation (GGA) proposed by Perdew, Burke, and Ernzerhof (PBE)[39], is selected to treat the exchange-correlation interactions between electrons. In order to reach the required accuracy in the electronic properties calculations of the studied systems, 500 eV is set as the cutoff energy in the plane-wave expanding, and a 21×21×1 k-points mesh by Monkhorst-Pack method is adopted in the Brillouin zone (BZ) sampling. Moreover, a vacuum layer of ~25 Å perpendicular to the monolayers is found to be enough to avoid the interactions between the periodic cells. All the atoms are fully relaxed until the total energy and the residue forces on each atom are converged to $10^{-4}$ eV and 0.01 eV/Å, respectively. In addition, compared with the PBE functional calculations, the Heyd-Scuseria-Ernzerhof (HSE06) functional with 25% Hartree-Fock exchange energy is selected to perform the hybrid-DFT calculations and evaluate more accurate band gaps of the porous structures.

## 3. Results and Discussion

### 3.1 Porous Structure and Stability

With regard to most 2D monolayer materials, the minimum energy structures are commonly known as both the planar and buckled ones [24]. For example, graphenen possesses the planar structure. However, silene forms the buckled structure. We hereby consider that the AlN monolayer structures are made of two different possible geometrical configuration unitswhich we call H and T types, as shown in Fig.1. The graphene-like AlN monolayer (g-AlN) is first studied and compared with the previous results, and is used for testing the correction and the reliability of our calculation. The lattice constant and the bond length are 3.128 Å and 1.806 Å for monolayer g-AlN, which are in accordance with the previous study values of 3.126 Å and 1.80 Å, respectively [40,41]. The lattice constants of bulk Wurtzite AlN (WZ-AlN) are also studied and the calculated results (a=b=3.11 Å, c=4.99 Å) are proved to be in agreement with previous experimental and theoretical work [28,42]. After geometry relaxation, the initial structure of H- and T-AlN monolayers become planar ones. Thereby, the AlN monolayer has only planar structures and is different from the porous graphene which has both planar and buckled structures.

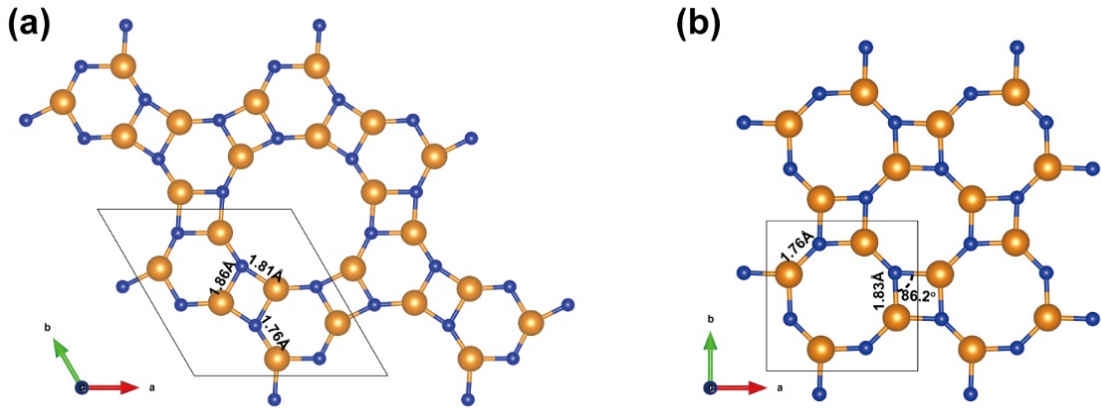

**Fig. 1** The geometry structures of two AlN monolayer. (a) H-AlN and (b) T-AlN phase. The blue and golden small balls are N and Al atoms, respectively.

From Fig. 1 (a) , we can see that the planar H-AlN structure can be described by the plane group 175.P6/M(C6H-1) with a lattice constant of 8.51 Å, with each hexagonal cell containing six Al atoms and six N atoms. The T-AlN structure can be described by the plane group 127.P4/MBM(D4H-5), and the lattice constant is 6.18 Å with each tetragonal cell containing four Al atoms and four N atoms. For clearer demonstration, we label the Al-N bond lengths and the angles in Fig. 1. For H-AlN structure, the unit cell is composed by two graphene-like honeycomb AlN units connected by a pair of Al-N bonds. The bond lengths in the two graphene-like honeycomb are 1.764 Å and 1.862 Å, while it is about 1.810 Å between the two graphene-like honeycombs. The T-AlN cell is an octagon structure and the unit cell is composed of one octagon and one square unit. The octagon unit has two alternate Al-N bonds and the bond lengths are 1.832 Å and 1.764 Å, while the value is 1.832 Å between the two octagons. Moreover, the theoretical surface areas of the H-AlN and T-AlN reach up to 2863 $m^2/g$ and 2615 $m^2/g$, which can be comparable with graphene 2630 $m^2/g$ [43,44].

To study the stability of the porous AlN monolayer, the formation energy is calculated firstly. Lower formation energy indicates that the single-layer is more stable, and it costs lower energy to synthesize.

The formation energy $E_f$ of AlN monolayer is defined as

$$E_f = \frac{E_{monolayer}}{m} - \frac{E_{bulk}}{n} \qquad (1)$$

Here, the $E_{bulk}$ is the energy of bulk WZ-AlN structures and $E_{monolayer}$ is the energy of the porous monolayer, respectively; the numbers $m$ and $n$ are the number of Al-N pairs in porous monolayer and WZ-AlN, respectively. According to Eq. (1), the formation energies of H-AlN, T-AlN, g-AlN are 1.47 eV, 1.32 eV, 0.93 eV, respectively. Based on the same definition of formation energy, the value is 1.52 eV (or 1.98eV) per Si dimer (or per Ge dimer) for silicene (or germanene), respectively [10, 45]. As silicene and germanene have been experimentally fabricated on some substrates, these porous structures may also be synthesized on proper substrates.

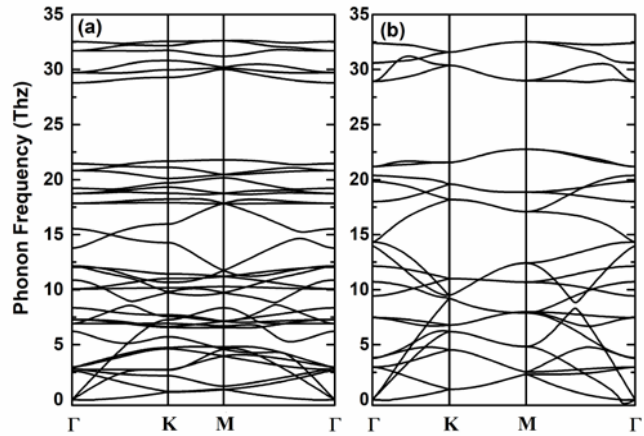

**Fig. 2** Phonon spectrum of H-AlN (a) and T-AlN (b) monolayer, respectively.

In order to investigate the structural stability of porous AlN monolayers, we calculate the phonon spectrum by density functional perturbation theory secondly and the results are shown in Fig. 2. It can be seen from the phonon dispersion that all the branches are basically positive frequencies in the whole Brillouin zone for both H-AlN and T-AlN monolayer. We cannot observe imaginary frequency part in the figure, which suggests that these two-porous monolayers are free-standing 2D materials with enough dynamic stability.

Thirdly, even if the dynamical stability of these H-AlN and T-AlN structures have been proved by the phonon spectrum analysis under T=0 K, the structure may be destroyed and transformed to other possible structure with high temperature. Hence, we have also examined the thermal stability with ab *initio* molecular dynamics simulations (AMDS). As for both porous monolayers, in order to avoid adverse effects of the periodic boundary condition, the AMDS use a 4×4×1 supercells which contain 192 and 128 atoms, respectively. The length of time-step is chosen as 2 fs and simulations with 3000 steps are executed in a canonical ensemble (NVT) at constant temperature T=300K and 1300K. The total potential energies per pair atom as a function of time-step and the structures at the end of simulation for H-AlN and T-AlN at different temperatures are represented as Fig. 3(a) (b) and (c) (d), respectively. The final structures of both porous monolayers simulation at the end of the AMDs are shown in Fig.3. Based on Fig.3, we can find that the total energies of the systems remain nearly constant, only showing a little deviation around the average energy during the simulation at 300K. At the end of the simulation, the structures show no significant differences as compared with the initial ones. Furthermore, a series of AMDS calculations are performed under different temperatures from 300 K to 1300 K with an interval of 200K. The total energy of the systems shows a small deviation around the average energy during the simulation process even at 1300K. The structure at the end of the simulation has a small wrinkle about 0.2~0.4 Å. The simulations indicate that the structure of H-AlN and T-AlN monolayer are not destroyed at 1300 K for 6 picoseconds. This also indicates that the H-AlN and T-AlN monolayer are stable at high temperature. All of these suggest that these porous monolayers have excellent structural stability and can be synthesized experimentally in the future at room temperature.

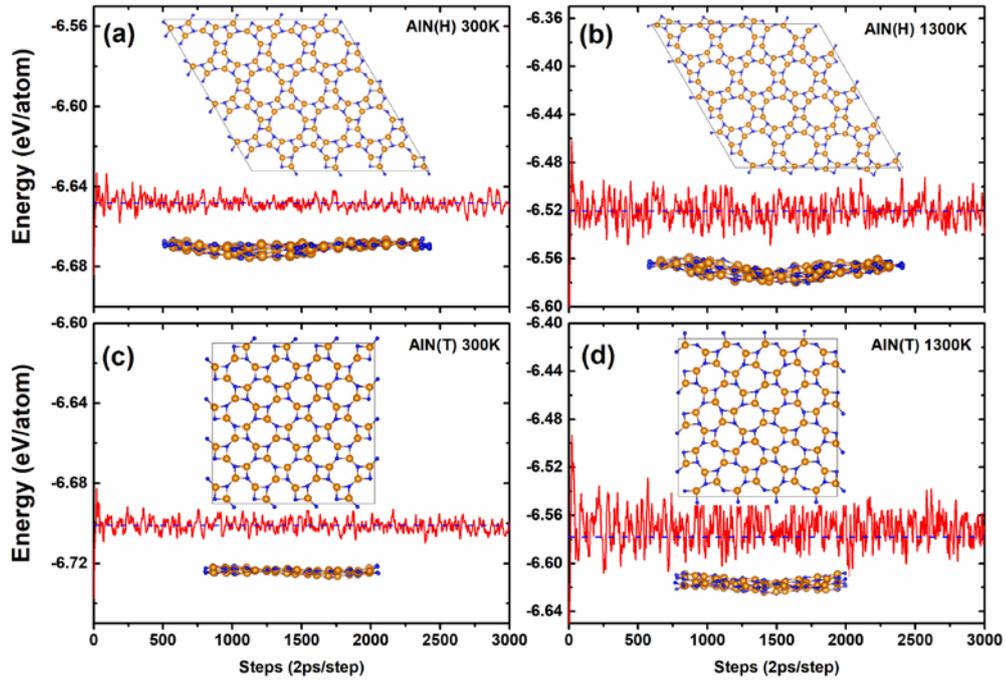

**Fig. 3** Relations of total energy and simulation time during ab initio molecular dynamics simulations: (a) H-AlN at 300K; (b) H-AlN at 1300K; (c) T-AlN at 300K; (d) T-AlN at 1300K, respectively. The AMDs are performed at T=300K and 1300K last for 6ps. The structures are also shown at the end of the AMDs.

## 3.2 Electronic Structures

The electronic characteristics of the porous AlN monolayer are studied by using GGA-PBE method. The band structures and the corresponding density of states are shown in Fig.4 and Fig.5. In order to compare it with the case of the nonporous, the g-AlN monolayer and WZ-AlN are also calculated, which are consistent with previous studies[42]. As shown in Fig.4, both H-AlN, T-AlN and g-AlN own an indirect band structure except the case of WZ-AlN. The indirect band gaps of the H-AlN and T-AlN monolayers are 2.89 eV and 2.86 eV, close to that 2.91 eV of g-AlN，indicating that these 2D monolayer materials may be useful for optoelectronic applications in the future.

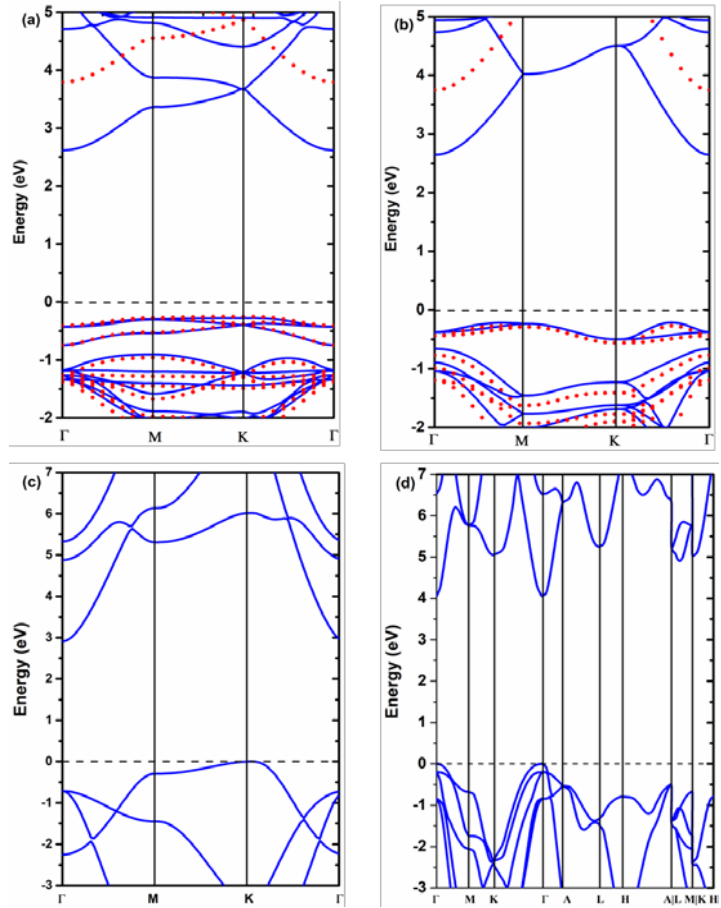

**Fig. 4** The band structures of (a) H-AlN and (b) T-AlN calculated with the GGA-PBE (blue) scheme; The red dashed line in band structures (a) and (b) obtained with the HSE06 scheme; the band structures of (c) g-AlN and (d) WZ-AlN calculated with the GGA-PBE (blue) scheme.

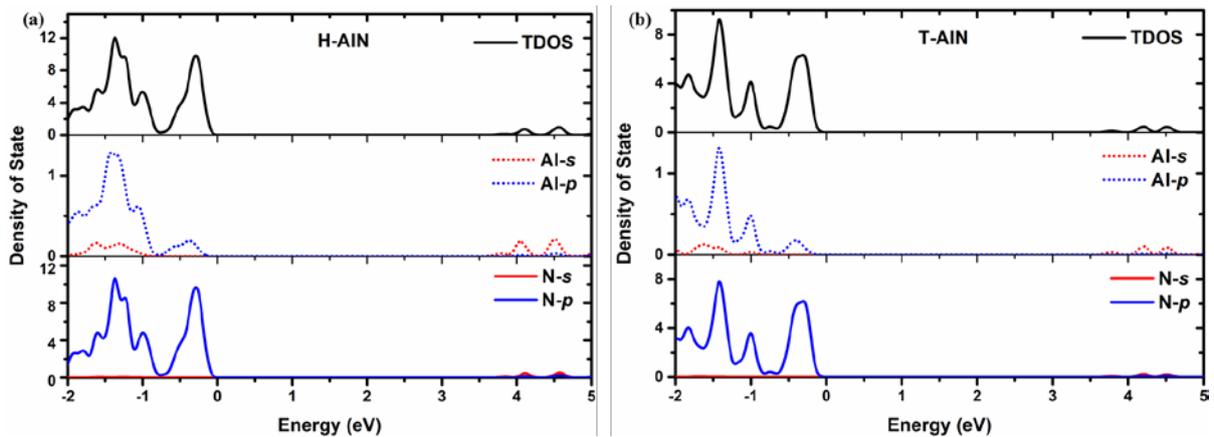

**Fig. 5** The density of states of (a) H-AlN and (b) T-AlN calculated with the GGA-PBE scheme.

In addition, the conduction band minimums (CBM) are all located at Γ point, but the valence band maximums (VBM) are located at different points. For H-AlN monolayer, the VBM is located on K point, while T-AlN at K-Γ line. As is shown in Fig.5, the density of states analysis for CBM shows that they all originate from the *s* orbitals of both Al and N atoms. Furthermore, for both H-AlN and T-AlN, the density of states suggest that the *p* orbitals of N atoms dominate the valence states near the Fermi level in comparison with that of Al atoms. This is similar to what is found in other allotropes of AlN and group-III nitrides. This can usually be attributed to the difference in the electronegativity of different atoms [47] and it is a common band characteristics in most honeycomb structure monolayer [11].

Moreover, due to the screened hybrid function, it can produce band gaps in much better agreement with experiment than GGA for group-III nitrides[48]. The hybrid-DFT method is also adopted to evaluate more accurate band gaps of the porous structures. Based on HSE06 functional with 25% Hartree-Fock exchange energy, the calculated band gap for WZ-AlN is 5.56eV, which is consistent with previous theoretically reported value[48]. This result agrees much better with previous experimental data (6.25eV) than the GGA value (4.1eV) [46; 48]. For H-AlN (T-AlN), the calculation gets an enlarged band gap 4.04 eV (4.01eV), as shown with the red dashed line in Fig.4 (a) and (b), respectively. However, despite the different magnitude of band gaps, the tendency of the obtained band structure doesn't change in comparison with the GGA-PBE results. In addition, we would like to point out that although the conventional DFT methods underestimate the band gap of semiconductors, it has been proved to be effective in predicting correct trends and physical mechanisms in previous works [7;49;50]. Therefore, for further related studies and analyses, the results obtained in this work are based on the conventional DFT methods with GGA-PBE scheme.

**3.3 Strain Effect on Tuning the Band Gap**

As a common and effective method for tuning the electronic properties of semiconductors, in-plane strain has been extensively used in theoretical calculations. Due to the novel structures, 2D materials can withstand much larger strains than the corresponding bulk crystals [48,51-54]. More precisely, increasing or decreasing the bond length as well as the confinement effects can modify the electronic structure. Thus, in order to obtain deeper insight about electronic properties, the biaxial compressive and tensile strains are applied to the AlN monolayers. The strain energy per unit cell is defined as follows,

$$E_s = E_T(\varepsilon) - E_T(\varepsilon = 0) \qquad (2)$$

Where $\varepsilon = \frac{a - a_0}{a_0} \times 100\%$ with $a_0$ being the lattice constant at the equilibrium state and $a$ the lattice constant at the strained state; $E_s$, $E_T(\varepsilon)$ and $E_T(\varepsilon = 0)$ are the strain energy, the total energy at applying a strain $\varepsilon$ and the total energy at the equilibrium states of one unit cell, respectively.

In order to obtain reliable results, the AlN monolayers are continuously applied biaxial strain with interval of 1%. The strain energies of H-AlN and T-AlN monolayer as a function of strain are presented in Fig. 6. It is found that the strain energies become nearly independent on the structure types and show similar tendency within the imposed strain range. In both cases, the strain energies increase continuously with increasing strain and without any suddenly drop in the process. If the strain energy falls suddenly, it usually indicates the material exhibit brittle fracture and local rupture. Structural optimization shows that the porous AlN monolayers have big change in band length and the shape of tetragonal cell, but they do not decompose in the simulations. If we release the strain and the deformation can disappear, the system may return to the initial situation without strain. The unique structure plays an important role in the whole process. Similar behavior is also found on nitrogene monolayer [55], which indicates elastic deformation under the strain. To further check the structural stability of the strained AlN monolayer, the phonon spectrum of these monolayers in the strained structures are calculated, which are shown in Fig. 7. As shown in Fig. 6, $\varepsilon_1$ and $\varepsilon_2$ are two critical strain obtained from the phonon spectrum calculations. We cannot observe imaginary frequency part for $\varepsilon_1 < \varepsilon < \varepsilon_2$, which suggests that the strained monolayers are dynamic stability in this range. The imaginary frequency part are appeared, when the strain is increased to $\varepsilon > \varepsilon_2$ or $\varepsilon < \varepsilon_1$. And in particular, the porous AlN monolayers are sensitive to the compressive strain. This behavior is similar with strained H-ScN monolayer, which can only remain stable by applying tensile strain according to the phonon dispersion curve [56].

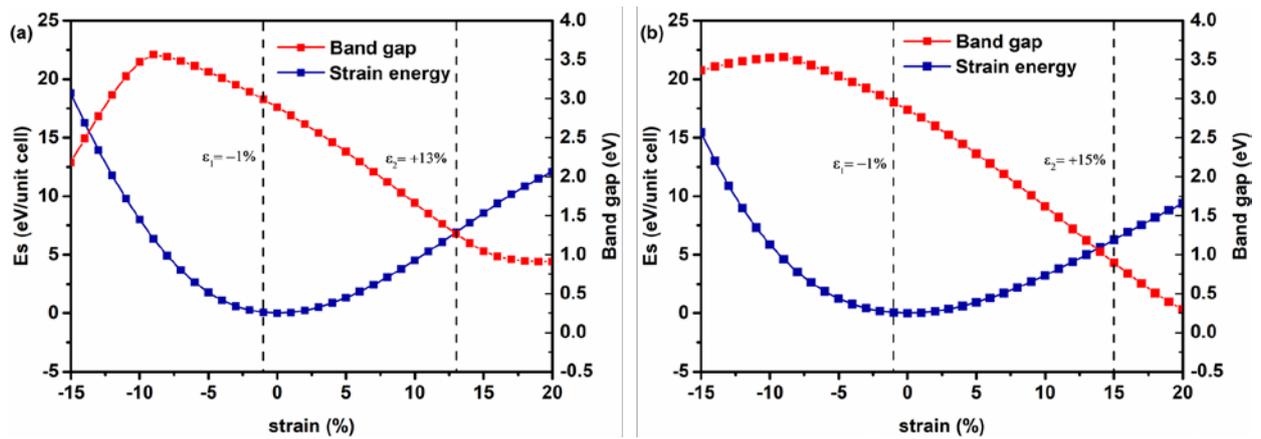

**Fig. 6** Strain energy and band gap of (a) H- and (b) T- AlN monolayer as a function of biaxial strains. $\varepsilon_1$ and $\varepsilon_2$ are two critical strain obtained from the phonon spectrum calculations.

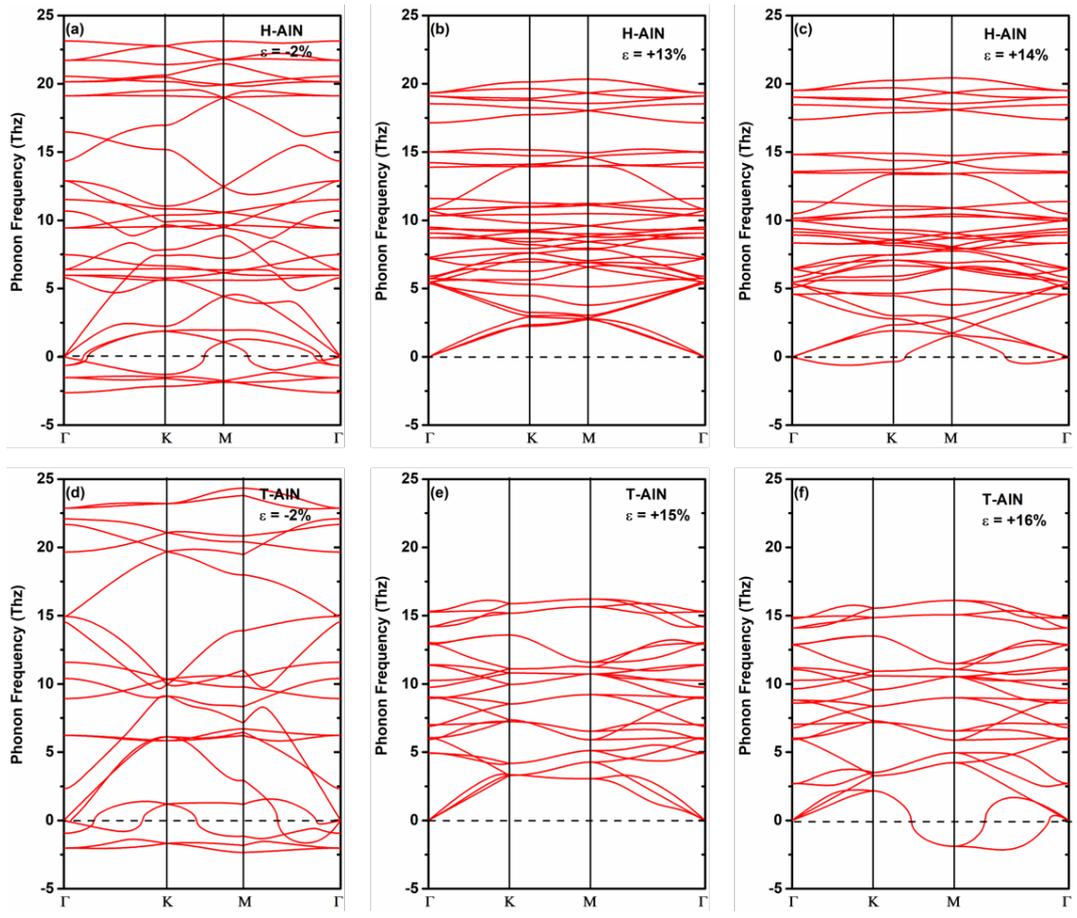

**Fig. 7** Phonon dispersion spectrum of strained (a) H- and (b) T- AlN monolayer.

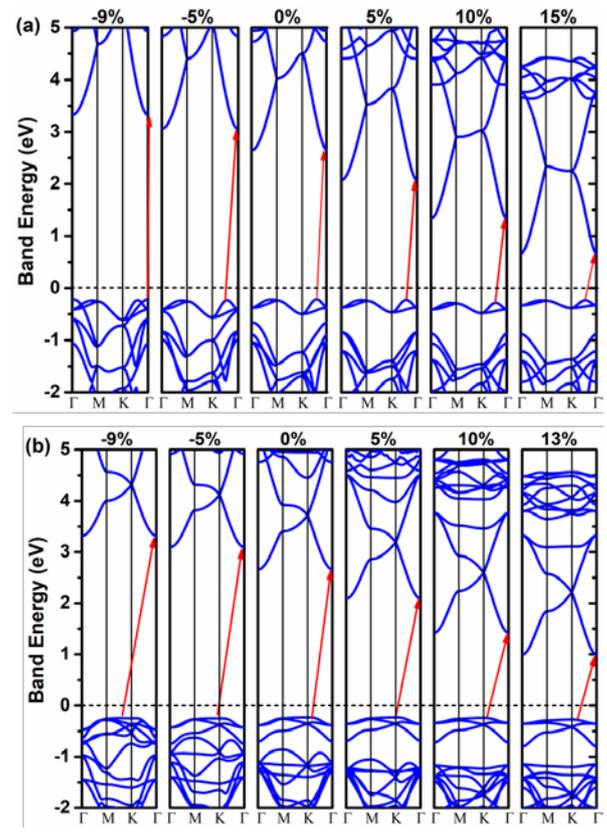

**Fig. 8** Band structures of (a) T- and (b) H-AlN monolayer under -20%, -15%, -9%, 0%, 5% and 10% biaxial strains.

To observe the effect of strain, Fig.6 exhibits the band gap of strained porous AlN monolayers. As can be seen from the figures, both H-AlN and T-AlN porous monolayer show various band gaps with different strains and the band gap can be tuned by applying strains. For tensile strain ($\varepsilon > 0$), the band gaps decrease monotonically while increasing the strain from -10% to +15%. These results are similar to the previous studies of the case of g-GaN, where the band gaps were reported to decrease linearly with the tensile increase[54]. The decrease in band gap can be attributed to the modification of orbitals under strain. Under tensile strains, the bond between neighbor atoms are stretched longer, the electron distribution and the orbital hybridization will be modified to adapt to the changes. This eventually leads to changes in band gaps. However, for compressive strain ($\varepsilon < 0$), the band gap increases and reaches the largest value of 3.56 (H-AlN) and 3.53 eV (T-AlN), while the compressive strain ($\varepsilon < 0$) increases from 0% to 9%. In conclusion, the band gap decreases under tensile strain while it increases to a maximum under moderate compressive strain and decreases again under extreme compressive strain. they show unique properties with strain compared with germanene, phosphorene, and so on [52; 57].

More interestingly, there is a direct-indirect band gap transition for T-AlN monolayer under strain. The electronic band structures under different strains for T-AlN, as well as H-AlN, are calculated and illustrated in Fig. 8. Our results indicate that H-AlN and T-AlN monolayers are indirect band gap semiconductors without strain. The CBM of porous monolayers are always located at Γ point. However, the VBM varies along the Γ-M-K-Γ path in the 2D Brillouin zone. As can be seen from the figures, an increase compressive strain from -0% to -9% leads to a direct-indirect transition of T-AlN. The VBM of T-AlN move to Γ point at -9% strain state, which results in a direct-indirect transition, as shown in Fig. 8.

Specially, while the compressive strain is further increased, the VBM shifts away from Γ point to K-Γ line and it transforms to an indirect band gap again. Although, the T-AlN monolayer becomes a direct band gap semiconductor at –9% strain state, the H-AlN monolayer remains indirect band gap semiconductor for all applied strain value.

This finding leads to the following discussions. First, a shortening of the bond length may play an important role in the direct-indirect transition. With compressive strain, the bond lengths between the neighbor atoms are shortened which leads to the increase of the repulsion between orbitals and exert different impacts on the in-plane and out-of-plane orbitals. This reorganizes the bands near the Fermi level and cause the VBM shifts[58]. The proximity orbital energies of N-2$p$ (8.4 eV) and Al-3$s$ (8.0 eV) is probably the second main factor[47]. Due to the short bond length and the proximity orbital energies in both allotropes, the N-2$p$ and Al-3$s$ form hybrid orbitals changing the valence states near the Fermi level.

## 4. Conclusion

In conclusion, two novel porous AlN monolayers are explored by using the density functional theory under the general gradient approximation expressed by the PBE functional. The stabilities are demonstrated by the phonon dispersion spectrum and the molecular dynamics simulation. Results show that these porous monolayers have excellent structural stability and can withstand temperatures as high as 1300K. Based on our findings, we perform a systematic investigation of electronic structures of these monolayers. Our results indicate that the H- and T- AlN monolayers are indirect-gap semiconductor with gaps of 2.89 eV and 2.86 eV at free strain case. Moreover, we reveal that the electronic structures can be modulated by strains. For T-AlN, there is an indirect-direct band gap transition with compressive strains. Due to its unique porous structures and high surface area, as well as the alterable bandgaps, these two monolayer materials may be useful for optoelectronic applications in the future.

## 5. Acknowledgments

This work was supported by the National Basic Research Program of China (Grant No. 2012CB921300), the National Natural Science Foundation of China (Grant No.11274280), the Foundation of Henan Educational committee (Grant No.20A140009) and Scientific Foundation of Henan University of Technology (Grant No.2018BS039). The calculations are supported by High Performance Computing Center of Zhengzhou University and Zhongyuan University of Technology and Zhengzhou Normal University.

## 6. Compliance with ethical standards

Conflict of interest    The authors declared that they have no conflict of interest to this work.